\parindent=0pt
\parskip=20pt
{\bf{}Simulation of finite state machines in a quantum computer}

Michael R. Dunlavey
\hfill\break
Pharsight Corporation
\hfill\break
200 Baker Ave.
\hfill\break
Concord, MA 01742
\hfill\break
mdunlave@pharsight.com

\baselineskip=20pt
{\bf{}ABSTRACT}

A construction is given for simulating any deterministic finite state
machine (FSM) on a quantum computer in a space-efficient manner.  By
constructing a superposition of input strings of lengths $K$ or less,
questions can be asked about the FSM, such as the inputs that reach
particular nodes, and the answers can be found using a search
algorithm.  This has implications for the eventual utility of quantum
computers for software validation.

{\bf{}1. Introduction}

An important question is: Can quantum parallelism be used to simulate
the execution of ordinary software on a very wide range of inputs.
If so, then it could be a useful tool for software validation.

A simpler question is: Can quantum parallelism simulate the execution
of a deterministic finite state machine (FSM) for all input strings
of length $K$ or less.  Various problems could be answered such as
which input strings cause the machine to halt at node $n$.

In broad outline, we make the FSM reversible by causing it to write
symbols to an output tape whenever it enters a node.  Then the
machine can be operated in reverse simply by reversing the roles of
input and output.  We assume the input and output alphabets are
binary, that the input and output strings are encoded as binary
numbers, and that the input and output tapes are represented as
quantum shift registers of sufficient length.

Figure 1 demonstrates a FSM set up in this way.  Arcs marked with
$\epsilon$ simply fall through without reading a symbol.

To answer questions, we can use the Grover search algorithm[1].  We
first place the input shift register into a superposition of all
possible inputs.  Then we evolve the FSM forward, invert the
amplitude of any quantum states in which the answer is true, and
devolve the FSM back to the initial condition.  Then we apply the
diffusion transform to enhance the probability of the marked states.
This is repeated the necessary number of times, and then an
observation is performed on the state.  This entire procedure is
repeated enough times to get a good statistical sample of the answer
states.

As an example of question answering, consider the FSM in Figure 2a.
The question is: Which input strings will cause the machine to halt
in node $2$?  Clearly the answer is all strings satisfying the regular
expression $0^*10^*1$,
but let us see how it would be done.

{\bf{}2. Encoding of input strings of length $\le{}K$}

To encode strings of various lengths in a binary input register of
fixed length, it is necessary to encode the length.  In this
construction, this is done by pre-pending a high-order $1$ bit.

Consider an input string $a$ of length $k\le{}K$ where
$a = \langle{}a_{k-1}, ... , a_0\rangle{}$
where $a_0$ is the first symbol to be read,
$a_1$ is the second, and so on.  The string is encoded
as a $K+1$-bit integer $I$:
$$I = (\sum_{i=0}^{k-1} a_i 2^i) + 2^k$$
The $2^k$ term appends a $1$-bit to the high-order end
of the integer $I$, and this will act as a terminating
marker.

{\bf{}3. Augmenting the FSM to handle encoded inputs}

In order to see if the FSM will stop in node $n$ when given input
$a$, we will need to augment node $n$ with an additional node $n'$
and network to detect the terminating $1$ bit followed by as many
remaining high-order $0$ bits as exist in the input register.
If after processing all $K+1$ bits of input the machine is in node
$n'$, then it would have stopped in node $n$ on the un-encoded
string.  This augmentation is shown in Figure 2b.  The original
machine stops in node $2$ just if the modified machine is found in
node $3$ after processing the entire input register.

{\bf{}4. Construction of the Reversible FSM}

A deterministic FSM is a tuple $\langle{}I, S, N, A\rangle{}$ where
\hfill\break
$I$ is the finite input vector defined over alphabet $S$,
\hfill\break
$S$ is the alphabet, in this case binary: $S = \{0,1\}$,
\hfill\break
$N$ is the set of nodes of the FSM, in this case numbered $0,1,...$,
\hfill\break
$A$ is the set of arcs of the FSM, $A \subseteq N \times S^+ \times N$
where $S^+ = S \cup \{\epsilon\}$ is the alphabet augmented with the
symbol $\epsilon$.  An $\epsilon$ in an arc indicates that the
arc "falls through" without reading an input symbol.

A reversible FSM is a tuple $\langle{}I, S, N, A_r, O\rangle{}$ in
which arcs of $A_r \subseteq N \times S^+ \times N \times S^+$ can
optionally write symbols to an output register $O$, where $O$ is a
vector over $S$.  An arc $a \in A_r$ is a tuple
$\langle{}n, s, m, t\rangle{}$
where $n$ is the source node, $s$ is the input symbol or $\epsilon$,
$m$ is the destination node, and $t$ is the output symbol or $\epsilon$.
If the input symbol $s$ is $\epsilon$, then $a$ can be the only arc
leaving $n$, and the arc is traversed without reading any input.
If the output symbol $t$ is $\epsilon$, then the arc $a$ does not
write an output symbol when it is traversed.

Define $\cdot{}n$ the set of {\sl{}upstream arcs} of node $n$ as
$\cdot{}n = \{\langle x,s,n,t\rangle  \in A_r\}$.

To construct $A_r$, first copy all arcs in $A$, giving $\epsilon$
as the output symbol:
$$\forall{}a \in A, a=\langle{}n, s, m\rangle{} \Rightarrow \langle{}n, s, m, \epsilon\rangle{} \in A_r$$
Then for each set of upstream arcs of
$n \in N$ such that $\vert{}\cdot{}n\vert{} > 2$ perform the following
replacement: where the upstream arcs are
$$\cdot{}n = \{\langle x_0,s_0,n,t_0\rangle , \langle{}x_1,s_1,n,t_1\rangle{}, ..., \langle{}x_k,s_k,n,t_k\rangle{}\}$$
they are replaced by:
$$\{\langle{}x_0,s_0,n',t_0\rangle{}, \langle{}x_1,s_1,n',t_1\rangle{}, ..., \langle{}x_k,s_k,n,t_k\rangle{}, \langle{}n',\epsilon{},n,\epsilon{}\rangle{}\}$$
where $n'$ is a newly created node.
Repeat this process until all nodes have $2$ or fewer upstream arcs.

At this point in the construction, all output symbols in $A_r$ are
$\epsilon$.  The next step is to supply binary output symbols.
For each $n\in{}N$ where
$$ \cdot{}n = \{\langle{}x_0,s_0,n,\epsilon{}\rangle{}, [\langle{}x_1,s_1,n,\epsilon{}\rangle{}]\}$$
replace these arcs by
$$      \{\langle{}x_0,s_0,n,0\rangle{}, [\langle{}x_1,s_1,n,1\rangle{}]\}$$
where the square brackets indicate the optional second arc.
Now the reversible FSM is complete.

To run the machine in reverse, run the machine 
$$\langle{}O,S,N,A_r^{-1},I\rangle{}$$
Note that $A_r^{-1}$ is a function, by construction.

{\bf{}5. Implement FSM in a Quantum Computer}

For the state machine, we will use a small register $N$ in which the
node $n$ is encoded as a binary number.

For input and output we will use registers $I$ and $O$, respectively,
which can be conditionally rotated in place depending on which node
the FSM is in.  The LSB of each register will be the point from which
bits are removed or inserted.  The conditional rotation of the input
register, denoted $U_i$, is a unitary transform on the bits of $N$
concatenated with those of $I$.  Similarly, the conditional rotation
of the output register $O$ is effected by unitary transform $U_o$,
operating upon the bits of $N$ and $O$.

Operation of the FSM is effected by $U_m$, the unitary transform of the
machine, operating on the bits of $N$, the LSB of $I$, and the LSB of $O$.

In the initial condition, the input register contains the input bit
string, or a superposition, shifted left by $1$ bit.  The output
register contains all zero bits, and $N$ contains the number $0$
representing the initial node.

To perform a step of computation, apply the transforms in the
sequence $U_i$, $U_m$, $U_o$.  If $N$ is in a reading state, $U_i$
shifts a data bit into the LSB.  $U_m$ changes the node from source
to destination.  If the source node was a reading node, the LSB of
the input register is cleared.  If the destination node is a writing
node, the LSB of the output register is set to the needed output.
Finally, if the destination node is a writing node, the $U_o$
transform does a left-rotate of the output register.  Since the
output register was initially empty, this rotates a zero bit into the
LSB.
Figure 3 shows the structure of the computation.

The machine can be reversed in an exactly symmetrical fashion by
applying transforms $U_o^{-1}$, $U_m^{-1}$, and $U_i^{-1}$ in
sequence.  In this case the roles of the output and input registers
are reversed.

{\bf{}Construction of $U_m$, $U_i$, and $U_o$}

The unitary machine transform, $U_m$, transforms the ket-vector
consisting of $\vert{}N, I_0, O_0\rangle$ where $I_0$ and $O_0$ are
the LSB of the input and output registers respectively, and $N$ is
the node register.  The transform is represented as any 1-1 onto
function of this ket vector that includes a specified table of pairs.
For example, the transform $U_m$ for the FSM in Figure 1 is specified as:

\vbox{
\baselineskip=12pt
\settabs 4 \columns
\hfill\break
\+&$	N I_0 O_0 \to N' I_0' O_0'$\cr
\+&\cr
\+&$	0, 0, 0 \to 1,  0,  0$\cr
\+&$	0, 1, 0 \to 4,  0,  0$\cr
\+&$	1, 0, 0 \to 2,  0,  0$\cr
\+&$	1, 1, 0 \to 3,  0,  0$\cr
\+&$	2, 0, 0 \to 6,  0,  1$\cr
\+&$	2, 1, 0 \to 6,  1,  1$\cr
\+&$	3, 0, 0 \to 5,  0,  1$\cr
\+&$	3, 1, 0 \to 5,  1,  1$\cr
\+&$	4, 0, 0 \to 5,  0,  0$\cr
\+&$	4, 1, 0 \to 5,  1,  0$\cr
\+&$	5, 0, 0 \to 6,  0,  0$\cr
\+&$	5, 1, 0 \to 6,  1,  0$\cr
\+&$	6, 0, 0 \to 0,  0,  0$\cr
\+&$	6, 1, 0 \to 0,  1,  0$\cr
}

It is assumed that this unitary transform and others could be
implemented in a physically realizable quantum computer by using
universal gates such as the Toffoli gate[2].

The input unitary transform $U_i$ is a $1-1$ onto function of
ket-vectors consisting of the $N$ register and the $I$ register,
$\vert{}N,I\rangle$ .  For nodes $n\in{}N$ that read input, the $I$
register is rotated to the right by one bit, where the LSB is
considered to be on the right.  For other nodes, the $I$ register is
unchanged.

The output unitary transform $U_o$ is a $1-1$ onto function of
ket-vectors consisting of the $N$ register and the $O$ register,
$\vert{}N,O\rangle$ .  For nodes $n\in{}N$ that, when entered, require
an output bit to be written, the $O$ register is rotated left by one
bit, where the LSB is considered to be on the right.  For other
nodes, the $O$ register is unchanged.

{\bf{}6. Unaddressed Questions}

The question remains: How many iterations of the sequence
$U_i$, $U_m$, $U_o$ are necessary to run the FSM to completion
on all possible inputs?  Is there a solution involving
some sort of cursor register[3], in which an isolated qubit
can be observed?
Related question: what happens if the machine runs too long?

In the search algorithm, how many iterations of the sequence
evolve-mark-devolve-diffusion are needed, given that the
number of marked states is not known in advance?

{\bf{}7. References}
\hfill\break
[1] L. K. Grover, A fast quantum mechanical algorithm for database
search, quant-ph/9605043
\hfill\break
[2] A. Barenco, et. al., Elementary gates for quantum computation,
quant-ph/9503016
\hfill\break
[3] C. P. Williams, and S. H. Clearwater, Explorations in Quantum Computing,
Springer-Verlag, N.Y., 1998

\end